\documentclass[12pt]{article}
\usepackage{graphicx}
\begin{document}

\vspace{6cm}

\begin{center}
{\large \bf Role of Quantum Optics in \\
Synthesizing Quantum Mechanics and Relativity *}

\vspace{3mm}

Y. S. Kim\footnote{electronic address: yskim@umd.edu}\\
Center for Fundamental Physics, University of Maryland,\\
College Park, Maryland 20742, U.S.A.

\end{center}

\vspace{1cm}

\abstract{Two-photon states produce enough symmetry needed for Dirac's
construction of the two-oscillator system which produces the Lie
algebra for the O(3,2) space-time symmetry.  This O(3,2) group
can be contracted to the inhomogeneous Lorentz group which,
according to Dirac, serves as the basic space-time symmetry for
quantum mechanics in the Lorentz-covariant world.  Since the
harmonic oscillator serves as the language of Heisenberg's
uncertainty relations, it is right to say that the symmetry of
the Lorentz-covariant world, with Einstein's $E = mc^2$, is
derivable from Heisenberg's uncertainty relations.  }

\vspace{8cm}
\noindent * based on an invited talk presented at the 26th
International Conference on Quantum Optics and Quantum Information
(Minsk, Belarus, May 2019).

\vspace{1cm}
\noindent For a pdf file with sharper images,
go to http://ysfine.com/yspapers/minsk19.pdf.

\newpage
\section{Introduction}
In 1963, Paul A. M. Dirac constructed the space-time symmetry of the deSitter
group $O(3,2)$~\cite{dir63}.  This deSitter group can be contracted to
the symmetry of the inhomogeneous Lorentz group which, according to Dirac,
is the fundamental equation for quantum mechanics in the Lorentz-covariant
world.

More recently, two-photon system became a prominent subject in physics.
The purpose of this paper is to point out that Dirac's $O(3,2)$ system can
be constructed from the two-photon systems of current interest.  In
1976~\cite{yuen76}, Yuen constructed the first formula for two-photon coherent
states known as squeezed states.  In 1986~\cite{yurke86}, Yurke, McCall, and
Klauder discussed two-photon interferometers exhibiting the U(1,1) and U(2) symmetries.
If we combine these two-photon operators into one algebraic system, we end up
with Dirac's $O(3,2)$ system.

Ever since Heisenberg declared his uncertainty relations in 1927, Paul
A. M. Dirac was interested in whether quantum mechanics is consistent with
Einstein's special relativity.
In 1927~\cite{dir27}, Dirac notes that the c-number time-energy uncertainty
relation causes a difficulty in making quantum mechanics Lorentz covariant.
In 1945~\cite{dir45}, Dirac uses a Gaussian form with the time variable to
construct a representation of the Lorentz group.  However, he does not
address the issue of the c-number nature of the time-energy uncertainty relation.

In 1949 paper in the special issue of the Reviews of Modern Physics in commemoration
of Einstein's 70th Birthday~\cite{dir49}, Dirac says that
the task of constructing relativistic quantum mechanics is constructing a
representation of the inhomogeneous Lorentz group.  In the same paper, Dirac
introduces the light-cone coordinate system telling the Lorentz boost is
a squeeze transformation.
In 1963, Dirac uses two coupled oscillators to construct the Lie algebra
for the $O(3,2)$ deSitter group.

Indeed, Dirac made his lifelong efforts to synthesize quantum mechanics and special
relativity.  One hundred years ago, Bohr was interested in the electron orbit
of the hydrogen atom.  Einstein was in worrying about how things look to
moving observers.  Dirac was interested in this problem, but it was a metaphysical
problem before 1960.

After 1950, the physics world started producing protons
moving with speed comparable with that of light.  In 1964, Gell-Mann produced
his quark model telling the proton, like the hydrogen atom, is a quantum bound state
of more fundamental particles called the quarks.  In 1969, Feynman noted that the
proton, when it moves with its speed close to that of light, appears as a collection
partons with their peculiar properties.

\begin{figure}%[thb]
\centerline{\includegraphics[width=16cm]{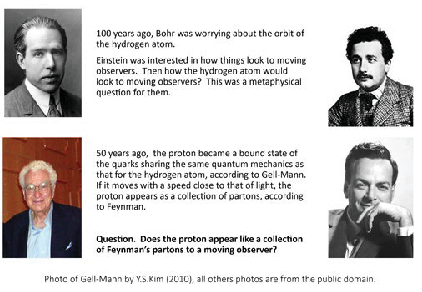}}
\caption{How the hydrogen atom look to moving observers?  Fifty
years later, this Bohr-Einstein issue  becomes the Gell-Mann-Feynman
issue. The issue is whether Feynman's partons are Gell-Mann's quarks
viewed by a moving observer.}\label{quapar11}
\end{figure}

\begin{figure}%[thb]
\centerline{\includegraphics[width=14cm]{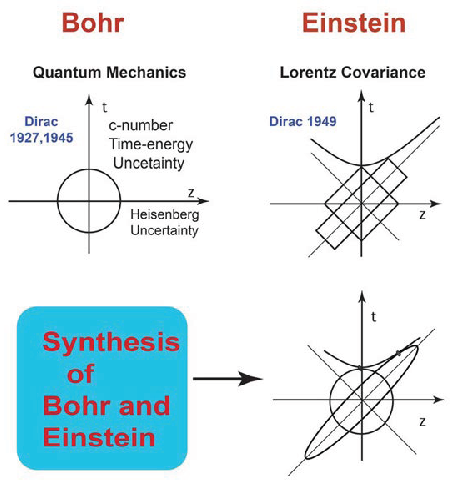}}
\caption{The hydrogen atom is a circle. The Lorentz boost is a squeeze
transformation.  If we combine them, the net effect is a squeezed
circle.}\label{synthes}
\end{figure}

\begin{figure}%[thb]
\centerline{\includegraphics[width=16cm]{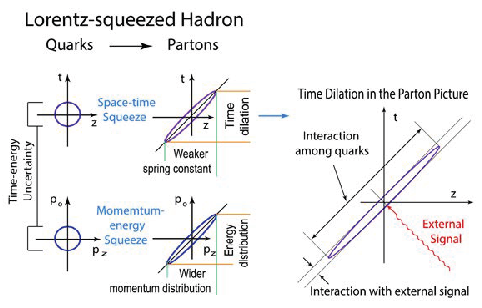}}
\caption{The crucial question is whether this squeezing effect can
be observed in laboratories.  This effects manifests itself through
the wide-spread parton momentum distribution, short interaction time,
partons as free light-like particles, as Feynman observed.
This figure is from Ref.~\cite{bkn15iop}.}\label{parton33}
\end{figure}

Thus the Bohr-Einstein issue became the Gell-Mann-Feynman issue, as specified in
Fig.~\ref{quapar11}.  The oscillator representation of Dirac~\cite{dir45} allows
us to use a circle in the longitudinal space-like and time-like coordinates.
The light-cone coordinate system Dirac introduced in 1949 tells the Lorentz boost
squeezed the oscillator circle into an ellipse as shown in Fig.~\ref{synthes}.
The question then is whether this effect of Lorentz squeeze can be observed
in the real world~\ref{parton33}.

In the papers written in and before 1949, Dirac was interested in combining
two scientific disciplines into one.  However, in 1963, by starting from
the harmonic oscillators which represent Heisenberg's uncertainty relations,
Dirac obtains the Lie algebra of the $O(3,2)$ deSitter group with ten generators,
which is the Lorentz group applicable to three space-like and two time-like
directions.

%----------------------------------------------------------------------
%----------------------------------------------------------------------

From his 1963 paper, we get a hint that this $O(3,2)$ group may be transformed
into the inhomogeneous Lorentz group, which is the fundamental symmetry group
for quantum mechanics in the Lorentz-covariant world according to
Dirac~\cite{dir49}.  This group also has ten generators.  Six of
them are for the Lorentz group and four of them are for space-time
translations.

As in the case of Ba{\c s}kal {\it et al.}~\cite{bkn19}, we  show in this
paper the inhomogeneous Lorentz group can be obtained from $O(3,2)$ via
the procedure of group contractions introduced first by In{\"o}n{\"u}
and Wigner in 1953~\cite{inonu53}.

In Sec.~\ref{2osc}, it is shown that two-photon states widely discussed in
the current literature produce the ten generators for Dirac's two-oscillators
system, which leads us to the Lie algebra of the $O(3,2)$ group. The
five-by-five matrices for the ten generators of this group are also given.
In Sec~\ref{dirac49}, the inhomogeneous Lorentz group is discussed.  It is
shown that this group can also be represented by five-by-five matrices.
Five-by-five expressions are given also for ten generators of this group.
In Sec.~\ref{contrac}, the $O(3,2)$ croup is contracted to the inhomogeneous
Lorentz group~\cite{wig39}.  The four-momentum operators generated in
this way corresponds to Einstein's $E = mc^{2}.$

%----------------------------------------------------------------------
%----------------------------------------------------------------------

%----------------------------------------------------------------------
%----------------------------------------------------------------------

\newpage

\section{Dirac's Two-oscillator System from Quantum \\Optics}\label{2osc}

In 1963, Dirac published a paper entitled ``A Remarkable Representation of
the (3 + 2) deSitter Group''~\cite{dir63}.  In this paper, he starts with
two oscillators with the following
step-up and step-down operators.
\begin{eqnarray}\label{105}
&{}& a_{1} = \frac{1}{\sqrt{2}}\left(x_{1} + iP_{1}\right), \qquad
a_{1}^{\dag} = \frac{1}{\sqrt{2}}\left(x_{1} - iP_{1}\right),   \nonumber\\[2ex]
&{}& a_{2} = \frac{1}{\sqrt{2}}\left(x_{2} + iP_{2}\right), \qquad
a_{2}^{\dag} = \frac{1}{\sqrt{2}}\left(x_{2} - iP_{2}\right) .
\end{eqnarray}
In terms of these operators, Heisenberg's uncertainty relations can be written as
\begin{equation}\label{107}
\left[a_{i}, a^{\dag}_{j}\right] = \delta_{ij} .
\end{equation}
with
\begin{equation}\label{109}
  x_{i} = \frac{1}{\sqrt{2}}\left(a_{i} + a^{\dag}_{i} \right), \qquad
  P_{i} = \frac{i}{\sqrt{2}}\left(a^{\dag}_{i} - a_{i} \right),
\end{equation}

With these sets of operators, Dirac constructed three generators of the
form
\begin{equation}\label{111}
 J_{1} = {1\over 2}\left(a^{\dag}_{1}a_{2} + a^{\dag}_{2}a_{1}\right) ,
 \quad
 J_{2} = {1\over 2i}\left(a^{\dag}_{1}a_{2} - a^{\dag}_{2}a_{1}\right),
 \quad
 J_{3} = {1\over 2}\left(a^{\dag}_{1}a_{1} - a^{\dag}_{2}a_{2} \right),
\end{equation}
and three more of the form
\begin{eqnarray}\label{115}
&{}& K_{1} = -{1\over 4}\left(a^{\dag}_{1}a^{\dag}_{1} + a_{1}a_{1} -
  a^{\dag}_{2}a^{\dag}_{2} - a_{2}a_{2}\right) ,   \nonumber \\[2ex]
&{}& K_{2} = +{i\over 4}\left(a^{\dag}_{1}a^{\dag}_{1} - a_{1}a_{1} +
  a^{\dag}_{2}a^{\dag}_{2} - a_{2}a_{2}\right) ,   \nonumber \\[2ex]
&{}& K_{3} = {1\over 2}\left(a^{\dag}_{1}a^{\dag}_{2} + a_{1}a_{2}\right) .
\end{eqnarray}
These $J_{i}$ and $K_{i}$ operators satisfy the commutation relations
\begin{equation}\label{lie11}
 [J_{i}, J_{j}] = i\epsilon _{ijk} J_{k} ,\quad
 [J_{i}, K_{j}] = i\epsilon_{ijk} K_{k} ,  \quad
[K_{i}, K_{j}] = -i\epsilon _{ijk} J_{k} .
\end{equation}
This set of commutation relations is identical to the Lie algebra of the
Lorentz group where, $J_{i}$ and $L_{i}$ are three rotation and three
boost generators respectively.

In addition, with the harmonic oscillators, Dirac constructed another set
consisting of
\begin{eqnarray}
&{}& Q_{1} = -{i\over 4}\left(a^{\dag}_{1}a^{\dag}_{1} - a_{1}a_{1} -
  a^{\dag}_{2}a^{\dag}_{2} + a_{2}a_{2} \right) ,   \nonumber \\[2ex]
&{}& Q_{2} = -{1\over 4}\left(a^{\dag}_{1}a^{\dag}_{1} + a_{1}a_{1} +
   a^{\dag}_{2}a^{\dag}_{2} + a_{2}a_{2} \right), \nonumber\\[3ex]
&{}& Q_{3} = \frac{i}{2}\left(a_{1}^{\dag}a_{2}^{\dag} - a_{1}a_{2}\right) .
\end{eqnarray}
They then satisfy the commutation relations
\begin{equation}\label{lie22}
[J_{i}, Q_{j}] = i\epsilon_{ijk} Q_{k} , \quad
[Q_{i}, Q_{j}] = -i\epsilon _{ijk} J_{k} .
\end{equation}
Together with the relation $[J_{i}, J_{j}] = i\epsilon _{ijk} J_{k}$ given
in Eq.(\ref{lie11}), $J_{i}$ and $Q_{i}$ and produce another set of closed
commutation relations for the generators of the Lorentz group.
Like $K_{i}$, the $Q_{i}$ operators act as boost generators.

In order to construct a closed set of commutation relations for all the
generators, Dirac introduced an additional operator
 \begin{equation}\label{303}
S_{0} = {1\over 2}\left(a^{\dag}_{1}a_{1} + a_{2}a^{\dag}_{2}\right) .
\qquad
\end{equation}
Then the commutation relations are
\begin{equation}\label{lie33}
  [K_{i}, Q_{j}] = -i\delta_{ij} S_{0} , \quad
[J_{i}, S_{0}] = 0 ,  \quad [K_{i}, S_{0}] =  -iQ_{i},\quad
[Q_{i}, S_{0}] = iK_{i} .
\end{equation}
Dirac then noted that these three sets of commutation relations given
in Eqs.~(\ref{lie11},\ref{lie22},\ref{lie33}) constitute the Lie
algebra for the group $O(3,2)$.  This group is applicable to the
five-dimensional space of $(x, y, z, t, s)$, where $x, y, z$ are
for three space-like coordinates, and $t$ and $s$ are for time-like
variables.  The generators are therefore five-by-five matrices.
These matrices are given in Table~\ref{tab11} and Table~\ref{tab22}.

%-------------------------------------------------------------------------------
\begin{table}%[thb]
\caption{Generators of the Lorentz group with three rotation and three boost
generators applicable to the five-dimensional space of, where $x, y, z$
are for space-like coordinates, $t$ and $s$ are for the time-like dimensions.
These generators are totally separated from the $s$ coordinate with zero
elements on their fifth row and fifth column.  The differential operators do
not contain the $s$ variable.}\label{tab11}
\vspace{0.5mm}
\begin{center}
\begin{tabular}{ccccccc}
\hline
\hline\\[-0.4ex]
\hspace{5mm}& Generators &\hspace{5mm} & Differential  &\hspace{8mm}& Matrix
\\[0.8ex]
\hline\\ [-1.0ex]
\hspace{5mm}&
$J_{1}$
&\hspace{5mm} &  $ -i\left(y\frac{\partial}{\partial z} - z\frac{\partial}{\partial y}\right) $
 & \hspace{8mm}&
$\pmatrix{0 & 0 & 0 & 0  & 0 \cr 0 & 0 & -i & 0 & 0 \cr
  0 & i & 0 & 0 & 0 \cr 0 & 0 & 0 & 0 & 0 \cr 0 & 0 & 0 & 0 & 0 }$
\\[6ex]
\hline\\ [-0.7ex]
\hspace{5mm}&
$J_{2} $
&\hspace{5mm} &
$  -i\left(z\frac{\partial}{\partial x} - x\frac{\partial}{\partial z}\right)$
                &  \hspace{8mm} &
$ \pmatrix{0 & 0 & i & 0 & 0 \cr 0 & 0 & 0 & 0 & 0 \cr
-i & 0 & 0 & 0 & 0 \cr 0 & 0 & 0 & 0 & 0 \cr 0 & 0 & 0 & 0 & 0}$
\\[6.0ex]
\hline\\ [-0.7ex]
\hspace{5mm}& $J_{3} $ &\hspace{5mm} &
$ -i\left(x\frac{\partial}{\partial y} - y\frac{\partial}{\partial x}\right)$
                  & \hspace{8mm}&
$ \pmatrix{0 & -i & 0 & 0 & 0 \cr i & 0 & 0 & 0 & 0 \cr
0 & 0 & 0 & 0 & 0 \cr 0 & 0 & 0 & 0 & 0 \cr\    0 & 0 & 0 & 0 & 0} $
\\[6.0ex]
\hline\\ [-0.7ex]
\hspace{5mm}& $K_{1} $ &\hspace{5mm} &
$ -i\left(x\frac{\partial}{\partial t} + t\frac{\partial}{\partial x}\right)$
                  & \hspace{8mm}&
$ \pmatrix{0 & 0 & 0 & i & 0 \cr 0 & 0 & 0 & 0 & 0 \cr
0 & 0 & 0 & 0 & 0 \cr i & 0 & 0 & 0 & 0 \cr    0 & 0 & 0 & 0 & 0} $
\\[6.0ex]
\hline\\ [-0.7ex]
\hspace{5mm}& $K_{2} $ &\hspace{5mm} &
$ -i\left(y\frac{\partial}{\partial t} + t\frac{\partial}{\partial y}\right)$
                  & \hspace{8mm}&
$ \pmatrix{0 & 0 & 0 & 0 & 0 \cr 0 & 0 & 0 & i & 0 \cr
0 & 0 & 0 & 0 & 0 \cr 0 & i & 0 & 0 & 0 \cr 0 & 0 & 0 & 0 & 0} $
\\[6.0ex]
\hline\\ [-0.7ex]
\hspace{5mm}& $K_{3} $ &\hspace{5mm} &
$ -i\left(z\frac{\partial}{\partial t} + t\frac{\partial}{\partial z}\right)$
                  & \hspace{8mm}&
$ \pmatrix{0 & 0 & 0 & 0 & 0 \cr 0 & 0 & 0 & 0 & 0 \cr
0 & 0 & 0 & i & 0 \cr 0 & 0 & i & 0 & 0 \cr    0 & 0 & 0 & 0 & 0} $
\\[6.0ex]
%----------------------------------------------
\hline
\hline\\[-0.4ex]
\end{tabular}
\end{center}
\end{table}
%----------------------------------------------------------------------------------

%--------------------------------------------------------------------------------------
\begin{table}%[thb]
\caption{Four additional generators for the $O(3,2)$.  Unlike those given in
Table~\ref{tab22}, the generators in this table have non-zero elements only in
the fifth row and the fifth column.  Every differential operator contains the $s$
variable. }\label{tab22}
\vspace{0.5mm}
\begin{center}
\begin{tabular}{ccccccc}
\hline
\hline\\[-0.4ex]
\hspace{5mm}& Generators &\hspace{5mm} & Differential  &\hspace{8mm}& Matrix
\\[0.8ex]
\hline\\ [-1.0ex]
\hspace{5mm}&
$Q_{1}$
&\hspace{5mm} &  $ -i\left(x\frac{\partial}{\partial s} + s\frac{\partial}{\partial x}\right) $
 & \hspace{8mm}&
$\pmatrix{0 & 0 & 0 & 0  & i \cr 0 & 0 & 0 & 0 & 0 \cr
  0 & 0 & 0 & 0 & 0 \cr 0 & 0 & 0 & 0 & 0 \cr i & 0 & 0 & 0 & 0 }$
\\[6ex]
\hline\\ [-0.7ex]
\hspace{5mm}&
$Q_{2} $
&\hspace{5mm} &
$  -i\left(y\frac{\partial}{\partial s} + s\frac{\partial}{\partial y}\right)$
                &  \hspace{8mm} &
$ \pmatrix{0 & 0 & 0 & 0 & 0 \cr 0 & 0 & 0 & 0 & i \cr
0 & 0 & 0 & 0 & 0 \cr 0 & 0 & 0 & 0 & 0 \cr 0 & i & 0 & 0 & 0}$
\\[6.0ex]
\hline\\ [-0.7ex]
\hspace{5mm}& $Q_{3} $ &\hspace{5mm} &
$ -i\left(z\frac{\partial}{\partial s} + s\frac{\partial}{\partial z}\right)$
                  & \hspace{8mm}&
$ \pmatrix{0 & 0 & 0 & 0 & 0 \cr 0 & 0 & 0 & 0 & 0 \cr
0 & 0 & 0 & 0 & i \cr 0 & 0 & 0 & 0 & 0 \cr  0 & 0 & i & 0 & 0} $
\\[6.0ex]
\hline\\ [-0.7ex]
\hspace{5mm}& $S_{0} $ &\hspace{5mm} &
$ -i\left(t\frac{\partial}{\partial s} - s\frac{\partial}{\partial t}\right)$
                  & \hspace{8mm}&
$ \pmatrix{0 & 0 & 0 & 0 & 0 \cr 0 & 0 & 0 & 0 & 0 \cr
0 & 0 & 0 & 0 & 0 \cr 0 & 0 & 0 & 0 & -i \cr  0 & 0 & 0 & i & 0} $
\\[6.0ex]
%--------------------------------------------------------------------------------
\hline
\hline\\[-0.4ex]
\end{tabular}
\end{center}
\end{table}
%----------------------------------------------------------------------------------

As Dirac stated in his paper~\cite{dir63}, it is indeed remarkable that
the two-oscillator system leads to the space-time symmetry of the (3 + 2)
deSitter group.  Even more remarkable is that this two-oscillator system
can be derived from quantum optics.  In optics, $a_{i}$ and $a^{\dag}_{i}$
act as the annihilation and creation operators.  For the two-photon system,
$i$ can be 1 or 2.

With these two sets of operators, it is possible to construct two-photon
states.  In 1976~\cite{yuen76}, Yuen considered the two-photon state
generated by
\begin{equation}\label{301}
 Q_{3} = \frac{i}{2}\left(a_{1}^{\dag}a_{2}^{\dag} - a_{1}a_{2}\right),
\end{equation}
which leads to the two-mode coherent state known as the ``squeezed state.''

Later, in 1986~\cite{yurke86}, Yurke {\it et al.} considered two-mode
interferometers.  In their study of two-mode states, they started with
$ Q_{3}$ given in Eq.(\ref{301}).  They then noted that, in one of
their interferometers, the following two additional operators are needed.
\begin{equation}\label{303}
K_{3} = {1\over 2}\left(a^{\dag}_{1}a^{\dag}_{2} + a_{1}a_{2}\right) ,
\qquad
S_{0} = {1\over 2}\left(a^{\dag}_{1}a_{1} + a_{2}a^{\dag}_{2}\right) .
\end{equation}
The three Hermitian operators from Eq.(\ref{301}) and Eq.(\ref{303})
satisfy the commutation relations
\begin{equation} \label{305}
\left[K_{3}, Q_{3}\right] = -iS_{0}, \qquad
\left[Q_{3}, S_{0}\right] = iK_{3}, \qquad
\left[S_{0}, K_{3}\right] = iQ_{3} .
\end{equation}
Yurke {\it et al.} called this device the $SU(1,1)$ interferometer.
The group $SU(1,1)$ is isomorphic to the $O(2,1)$ group or the
Lorentz group applicable to two space-like and one time-like dimensions.

In addition, in the same paper~\cite{yurke86}, Yurke {\it et al.}
discussed the possibility of constructing another interferometer exhibiting
the symmetry generated by
\begin{equation}\label{307}
 J_{1} = {1\over 2}\left(a^{\dag}_{1}a_{2} + a^{\dag}_{2}a_{1}\right) , \quad
 J_{2} = {1\over 2i}\left(a^{\dag}_{1}a_{2} - a^{\dag}_{2}a_{1}\right), \quad
 J_{3} = {1\over 2}\left(a^{\dag}_{1}a_{1} - a^{\dag}_{2}a_{2} \right).
\end{equation}
These generators satisfy the closed set of commutation relations
$
\left[J_{i}, J_{j}\right] = i\epsilon_{ijk} J_{k} ,
$
given in Eq.(\ref{lie11}).  This is the Lie algebra for the three-dimensional
rotation group.  Yurke {\it et al.} called this optical device the $SU(2)$
interferometer.

We are then led to ask whether it is possible to construct a closed set
of commutation relations with the six Hermitian operators from
Eq.(\ref{305}) and Eq.~(\ref{307}). It is not possible.  We have to
add four additional operators, namely
\begin{eqnarray}\label{311}
&{}& K_{1} = -{1\over 4}\left(a^{\dag}_{1}a^{\dag}_{1} + a_{1}a_{1} -
  a^{\dag}_{2}a^{\dag}_{2} - a_{2}a_{2}\right) ,    \nonumber \\[1ex]
&{}&   K_{2} = +{i\over 4}\left(a^{\dag}_{1}a^{\dag}_{1} - a_{1}a_{1} +
  a^{\dag}_{2}a^{\dag}_{2} - a_{2}a_{2}\right) ,   \nonumber \\[1ex]
&{}& Q_{1} = -{i\over 4}\left(a^{\dag}_{1}a^{\dag}_{1} - a_{1}a_{1} -
  a^{\dag}_{2}a^{\dag}_{2} + a_{2}a_{2} \right) ,    \nonumber \\[1ex]
&{}& Q_{2} = -{1\over 4}\left(a^{\dag}_{1}a^{\dag}_{1} + a_{1}a_{1} +
   a^{\dag}_{2}a^{\dag}_{2} + a_{2}a_{2} \right) .
\end{eqnarray}
There are now ten operators.  They are precisely those ten Dirac
constructed in his paper of 1963~\cite{dir63}.

It is indeed remarkable that Dirac's $O(3,2)$ algebra is produced by
modern optics. This algebra produces  the Lorentz group applicable
to three space-like and two time-like dimensions.

%--------------------------------------------------------------------------
\begin{table}%[thb]
\caption{Generators of translations in the four-dimensional Minkowski
space.  We are eventually interested in converting the four generators in
the $O(3,2)$ group in Table~\ref{tab22}
into the four translation generators.}\label{tab33}
\vspace{0.5mm}
\begin{center}
\begin{tabular}{ccccccc}
\hline
\hline\\[-0.4ex]
\hspace{5mm}& Generators &\hspace{5mm} & Differential  &\hspace{8mm}& Matrix
\\[0.8ex]
\hline\\ [-1.0ex]
\hspace{5mm}&
$Q_{1}\rightarrow P_{1}$
&\hspace{5mm} &  $ -i\frac{\partial}{\partial x} $
 & \hspace{8mm}&
$\pmatrix{0 & 0 & 0 & 0  & i \cr 0 & 0 & 0 & 0 & 0 \cr
  0 & 0 & 0 & 0 & 0 \cr 0 & 0 & 0 & 0 & 0 \cr 0 & 0 & 0 & 0 & 0 }$
\\[6ex]
\hline\\ [-0.7ex]
\hspace{5mm}&
$Q_{2}\rightarrow P_{2} $
&\hspace{5mm} &
$  -i\frac{\partial}{\partial y}$
                &  \hspace{8mm} &
$ \pmatrix{0 & 0 & 0 & 0 & 0 \cr 0 & 0 & 0 & 0 & i \cr
0 & 0 & 0 & 0 & 0 \cr 0 & 0 & 0 & 0 & 0 \cr 0 & 0 & 0 & 0 & 0}$
\\[6.0ex]
\hline\\ [-0.7ex]
\hspace{5mm}&
$Q_{3} \rightarrow P_{3} $ &\hspace{5mm} &
$ -i\frac{\partial}{\partial z}$
                  & \hspace{8mm}&
$ \pmatrix{0 & 0 & 0 & 0 & 0 \cr 0 & 0 & 0 & 0 & 0 \cr
0 & 0 & 0 & 0 & i \cr 0 & 0 & 0 & 0 & 0 \cr  0 & 0 & 0 & 0 & 0} $
\\[6.0ex]
\hline\\ [-0.7ex]
\hspace{5mm}&
$S_{0}\rightarrow P_{0} $ &\hspace{5mm} &
$ i \frac{\partial}{\partial t}$
                  & \hspace{8mm}&
$ \pmatrix{0 & 0 & 0 & 0 & 0 \cr 0 & 0 & 0 & 0 & 0 \cr
0 & 0 & 0 & 0 & 0 \cr 0 & 0 & 0 & 0 & -i \cr  0 & 0 & 0 & 0 & 0} $
\\[6.0ex]

%----------------------------------------------
\hline
\hline\\[-0.4ex]
\end{tabular}
\end{center}
\end{table}
%----------------------------------------------------------------------------------

\section{Dirac's Forms of Relativistic Dynamics}\label{dirac49}
In 1949~\cite{dir49}, Paul A. M. Dirac published a paper entitled ``Forms
of Relativistic Dynamics,'' where he stated that the construction of
relativistic dynamics is to find a representation of the inhomogeneous
Lorentz group~\cite{wig39}.  This group is generated by three rotation
generators, three boost generators, and four translation generators.
If we use $J_{i}$ and $K_{i}$ for the rotation and boost generators
respectively, and $P_{i}$ and $P_{0}$ for the  four momentum generators,
they satisfy the following set of commutation relations.
\begin{equation}  \label{501}
 \left[J_{i}, J_{j}\right] = i\epsilon_{ijk} J_{k}, \quad
 \left[J_{i}, K_{j}\right] = i\epsilon_{ijk} K_{k}, \quad
   \left[K_{i}, K_{k}\right] = -i\epsilon_{ijk} J_{k},
\end{equation}
and
\begin{eqnarray}\label{502}
&{}& \left[P_{i}, J_{k}\right] = -i\epsilon_{ijk} J_{k}, \quad
\left[P_{i}, K_{k}\right] = -i\epsilon_{ijk} K_{k}, \nonumber \\[1.0ex]
&{}& \left[P_{i}, P_{i}\right] = 0, \quad
\left[P_{i}, P_{0}\right] = 0, \quad
\left[P_{0}, J_{i}\right] = \left[P_{0}, K_{i}\right] = 0   .
\end{eqnarray}
There are ten generators, as in the case of the $O(3,2)$ group. Among
them, the rotation and translation generators are Hermitian and correspond
to observable dynamical variables, while the boost operators do not.

As far as the Lorentz transformations are concerned, we can use
four-by-four matrices.  However, if we augment translations, we have
to use the transformation of the type
\begin{equation}
   \pmatrix{x + x' \cr y + y' \cr z + z' \cr t + t' \cr 1} =
    \pmatrix{ 1 & 0 & 0 & 0 & x' \cr    0 & 1 & 0 & 0 & y' \cr
    0 & 0 & 1 & 0 & z' \cr    0 & 0 & 0 & 1 & t'
    \cr 0 & 0 & 0 & 0 & 1 } \pmatrix{x \cr y \cr z \cr t \cr 1}.
\end{equation}
This five-by-five matrix is constructed from translation generators
from~Table~\ref{tab33} according to
\begin{equation}
\exp\left\{-i\left(x'P_{1} + y'P_{2} + z'P_{3} + t'P_{0} \right)\right\} .
\end{equation}

In this five-by-five representation, the $J_{i}$ and $K_{i}$ generators can
be written as the five-by-five matrices given in Table~\ref{tab11}.  The four
translation generators are given in Table~\ref{tab33}.  There are ten
generators, and the satisfy the Lie algebra of the inhomogeneous Lorentz given
in Eq.(\ref{501}) and Eq.(\ref{502}).  Table~\ref{tab33} indicates also that
these translation operators can be obtained from $Q_{i}$ and $S_{0}$ of the
$O(3,2)$ group discussed in Sec.~\ref{2osc}.  We shall see how this happens
in Sec.~\ref{contrac}.

\section{Contraction of O(3,2) to the Inhomogeneous \\
Lorentz Group}\label{contrac}

We are interested in transforming the group $O(3,2)$ into the indigenous
Lorentz group by contracting the $s$ coordinate according to the group
contraction procedure introduced  first by E. In{\"o}n{\"u} and E. P.
Wigner~\cite{inonu53}.  This procedure is applicable to contracting
the three-dimensional rotation group into a two-dimensional Euclidean group.
It is like the surface of the earth is flat for a limited area.  The same
In{\"o}n{\"u}-Wigner process can be used for contracting the Lorentz group
into the group of three-dimensional rotations and three translations.

The idea of the present section is to use the same contraction procedure
in order to convert $O(3,2)$ to the Lorentz group applicable to the
four-dimensional Minkowski space and four translations.

For this purpose, we cab make the $s$ coordinate continuously smaller to
zero in the limit:
\begin{equation}\label{533}
\pmatrix{1 & 0 & 0 & 0 & 0 \cr 0 & 1 & 0 & 0 & 0 \cr 0 & 0 & 1 & 0 & 0 \cr
0 & 0 & 0 & 1 & 0 \cr 0 & 0  & 0 & 0 & \epsilon }
\pmatrix{x \cr y \cr z \cr t \cr  s } \quad \rightarrow \quad
\pmatrix{x \cr y \cr x \cr t \cr 0} ,
\end{equation}
The contracted vector with $s = 0$ remains invariant under the inverse
transformation.

Let us use the notation  $C(\epsilon)$ for the five-by-five matrix given in
Eq.(\ref{533}).  This matrix commutes with $J_{i}$ and $K_{i}$ given in
Table~\ref{tab11}.  As for those in Table~\ref{tab22}, the same
transformation on the matrix $Q_{1}$ is
\begin{equation}
C~Q_{1}~C^{-1} = \pmatrix{0 & 0 & 0 & 0 & i/\epsilon \cr 0 & 0 & 0 & 0 & 0 \cr
0 & 0 & 0 & 0 & 0 \cr 0 & 0 & 0 & 0 & 0  \cr i~\epsilon  & 0 & 0 & 0 & 0} ,
\end{equation}
which in, in the limit of small $\epsilon$, becomes
\begin{equation}
C~Q_{1}~C^{-1} = \pmatrix{0 & 0 & 0 & 0 & i/\epsilon \cr 0 & 0 & 0 & 0 & 0 \cr
0 & 0 & 0 & 0 & 0 \cr 0 & 0 & 0 & 0 & 0  \cr 0 & 0 & 0 & 0 & 0} ,
\end{equation}
This matrix had only one non-zero element.  Thus inverse of this
transformation leads to
\begin{equation}
 \epsilon~C~Q_{1}~C^{-1} = \pmatrix{0 & 0 & 0 & 0 &  i \cr 0 & 0 & 0 & 0 & 0 \cr
0 & 0 & 0 & 0 & 0 \cr 0 & 0 & 0 & 0 & 0  \cr 0 & 0 & 0 & 0 & 0} ,
\end{equation}
This is precisely the matrix for the translation operator given in
Table~\ref{tab33}.  We can carry out similar procedures for other items
in the same table to complete the contraction of Dirac's $O(3,2)$ into
the inhomogeneous Lorentz group.

Special relativity and quantum mechanics have been and still are two
major physical theories formulated during the past century. It is
gratifying to note that these theories are derivable from the same
mathematical base, namely the mathematics of two coupled oscillators.

\section{Acknowledgments}

\begin{figure}%[thb]
\centerline{\includegraphics[width=16cm]{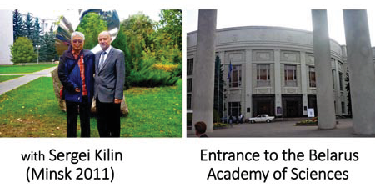}}
\caption{My photo with Sergei Kilin.  He has been organizing international
conference on quantum optics for many years, some at his home base of
Minsk and some others at various places in Europe. }\label{kilin}
%--------------------------------------------------------------------
\vspace{30mm}
\centerline{\includegraphics[width=16cm]{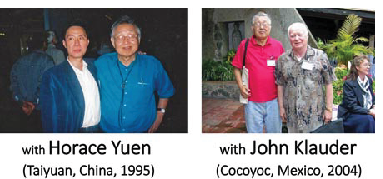}}
\caption{My photos with Horace Yuen and John Klauder.  This report starts
from their papers on two photons ~\cite{yuen76,yurke86}.}\label{yuen}
\end{figure}

This report is based on an invited talk presented at the 26th
International Conference on Quantum Optics and Quantum Information
(Minsk, Belarus, May 2019).  This conference was organized by Professor
Sergei Kilin of the Belarus Academy of Sciences.  I attended a number
of earlier conferences he organized in Minsk and other countries.  In
2011, I was fortunate enough to have a photo with him at the backyard
of the Belarus Academy of Science in Minsk as shown in Fig.~\ref{kilin}.
This report is, in part, based on the paper I published with
Sibel Ba{\c s}kal and Marilyn Noz~\cite{bkn19}.  I would like to thank
them for their many years of collaboration.

My professional background is in high-energy physics, and I am proud of
producing the result summarized in the first three figures given in
this paper.  They are based on the Lorentz group, which is the mathematics
of the Lorentz group.  While reading Horace Yuen's paper on two-photon coherent
states~\cite{yuen76}, I realized that the Lorentz group could be the
basic language for modern optics.  Sensing my new interest, John Klauder
sent me six copied of the reprint of the paper he published with Yurke and
McCall~\cite{yurke86}.  These two quantum optics papers, together with
Dirac's 1963 paper~\cite{dir63} serve as the building blocks of the present
paper.  I am very happy to show my photos with Yuen and Klauder in
Fig.~\ref{yuen}.

Dirac's 1963 paper on the $O(3,2)$ group is not widely known, and I am
the only one talking about this paper constantly.  There is a good
reason.  I met Dirac in October of 1962 after he completed this
article~\cite{dir63}.

%----------------------------------------------------------------------
%----------------------------------------------------------------------

Dirac did not talk to too many people.  How did I meet him?  I finished
my PhD degree at Princeton in 1961 and stayed there for one more year
as a post-doc before becoming an assistant professor at the Univ. of
Maryland in 1962.  At that time, the chairman of the physics department
was John S. Toll, and he was an ambitious man.  He invited Dirac in
October of 1962 for one week.  Since I was the youngest faculty member
in his department, Toll assigned me as Dirac's personal assistant.

At that time, I was publishing my papers acceptable to the American
physics community.  I had to write my papers starting from the premise
that the physics starts from singularities in the two-dimensional
complex plane, but I knew that I was writing useless (if not wrong)
papers.  In 2004, I wrote a report telling how much I disliked my
the papers written before 1965~\cite{dashen04}.

\begin{figure}%[thb]
\centerline{\includegraphics[width=16cm]{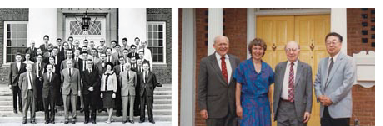}}
\caption{Physics faculty photo of 1963 and my photo with John Tall,
Mrs. Toll, and Eugene Wigner at the Chancellor's Mansion of the
University of Maryland (1986). }\label{tollwig}
\vspace{20mm}
\end{figure}

Dirac indeed taught me how to do physics: synthesize quantum
mechanics and relativity.   I was like Nicodemus meeting Jesus (story
from the Gospel of John in the New Testament).  I was born again.
Figure~\ref{tollwig} shows the physics faculty photo taken in the spring
of 1962. John Toll is in the middle of the first row.  I am the youngest
person in this figure.  By 1986, Toll was the Chancellor of the
University Maryland System.  He became very happy whenever Eugene
Wigner of Princeton came to the University of Maryland at my invitation.
Academic life as a faculty member is not always easy.  John Toll was
on my side whenever I needed him.  I am eternally grateful to him.

%----------------------------------------------------------------------
%----------------------------------------------------------------------

\end{document}